\documentstyle[epsf]{l-aa}

\newcommand{\p}[2]{\frac{\partial#1}{\partial#2}}
\newcommand{\up}[3]{ \left\{ \begin{array}{ll} #1 & \mbox{if $#2 \geq 0$} \\
							#3 & \mbox {otherwise}
							\end{array}
							\right.}
\newcommand{\av}[2]{{\rm av}\left(#1,#2\right)}
\newcommand{\rf}{\par\noindent\hangindent 20pt}
\newcommand{\AAP}[2]{A\&A {#1}, #2}

\newcommand{\AJ}[2]{AJ {#1}, #2}

\newcommand{\APJ}[2]{ApJ {#1}, #2}
\newcommand{\APJL}[2]{ApJ {#1}, L#2}
\newcommand{\APJS}[2]{ApJS {#1}, #2}

\newcommand{\MN}[2]{MNRAS {#1}, #2}

\begin{document}
\thesaurus{12(02.08.1; 02.09.1; 09.10.1; 08.05.1)}
\title{Numerical simulations of the Kelvin-Helmholtz instability in
radiatively cooled jets}
\author{T. P. Downes 
\and T. P. Ray}
\offprints{T. P. Downes}
\institute{Dublin Institute for Advanced Studies, 5 Merrion Square, Dublin
2, Ireland}
\date{Received date ;accepted date}
\maketitle

\begin{abstract}
We present the results of simulations of the development of the 
Kelvin-Helmholtz (KH) instability in a cooled, slab symmetric system.  The 
parameters were chosen to approximate the physical conditions typically found 
in jets from young stellar objects (YSOs).  The effect of different methods of 
maintaining the initial equilibrium were examined for varying density.  In 
addition, the effect of adjusting the width of the shear layer between the jet 
and ambient material was studied and found not to have significant long-term 
effects on the development of the instability.

We find that, in general, cooling acts to 
\begin{itemize}
\item increase the level of mixing between jet and ambient material through
the `breaking' of KH induced waves on the surface of the jet
\item increase the amount of momentum transferred from jet material to
ambient material
\item increase the time taken for shocks to develop in the flow
\item reduce the strength of these shocks
\item reduce the rate of decollimation of momentum flux
\end{itemize}
The first and second of these results appear to contradict the 
conclusions of Rossi et al.\ (1997) who carried out a similar study to ours
but in cylindrical symmetry.  It is found, however, that the differences 
between slab and cylindrical symmetry, while insignificant in the linear 
regime, explain the apparent discrepancy between our results and those of 
Rossi et al.\ (1997) in the non-linear regime.

\end{abstract}

\section{Introduction}

Of fundamental importance to the propagation of astrophysical jets is the 
question of their stability. Jet flows are known to vary over time and 
certainly in the case of both jets from active galactic nuclei (AGN), e.g., 
the M~87 jet (Harris et al.\ 1997), and from young stellar objects (YSOs), 
e.g.\ the L1551~IRS~5 jet (Neckel \& Staude 1987), this variability can 
often be traced to the source itself. Moreover, the dynamical interaction 
between faster jet fluid catching up with slower jet material can 
explain many of the features that we see for example in YSO jets (Raga 
et al.\ 1990; Stone \& Norman 1993).
At the same time, a number of fluid dynamic instabilities can occur even when 
the jet flow is, at least initially, very steady. For example, in purely 
hydrodynamic flows, Rayleigh-Taylor and Vishniac instabilities may be 
important especially at the head of YSO jets (Blondin et al.\ 1990). 
An additional concern is whether or not Kelvin-Helmholtz (KH) instabilities 
are important. Certainly KH instabilities are a viable explanation for some 
morphological features observed in AGN jets: for example the well-known 
sinusoidal appearance of the 3C449 jets has been successfully modeled by the 
non-linear growth of the helical twisting mode (Hardee et al.\ 1994). 

Until recently, most treatments of KH instabilities in jets, both of 
an analytical and numerical nature (e.g.\ Birkinshaw 1991; Bodo et al.\ 1994) 
have ignored the effects of radiative losses. In the case of AGN jets 
such an approach is justified since radiative losses can, to a first 
approximation be ignored. The same cannot be said of their smaller scale
counterparts i.e.\ jets from YSOs (e.g.\ Edwards et al.\ 1993; Ray 1996) 
where such energy losses have major effects on the development of the flow 
(Blondin et al.\ 1990; Stone \& Norman 1994 and references therein).
B\"uhrke et al.\ (1988) were the first to suggest that the knots seen in 
YSO jets might be due to the growth of KH pinching modes. Although it is 
now generally agreed, especially as a result of Hubble Space telescope 
imaging (e.g.\ Ray et al.\ 1997; Reipurth et al.\ 1997), that such knots are 
probably produced by episodic variations in the outflow from the source, it 
is nevertheless interesting to see what effects, if any, the growth of KH 
modes might have on the development of YSO jets. Effects caused by KH 
instabilities could be present in addition to those caused by ``pulsing''.
Another interesting question is whether the growth of KH modes might provide 
a mechanism through which momentum could be transferred to the jet's 
surroundings and thus accelerate ambient gas. Such momentum transfer 
would be in addition to any acceleration provided by the so-called ``prompt''
entrainment of ambient material near the head of the jet (Padman et al.\ 1997).

Recently a number of researchers have investigated the growth of KH modes in 
radiatively cooled slab jets. Hardee \& Stone (1997), in a linear stability 
analysis of such jets, solved the KH dispersion relationship numerically for 
a broad range of perturbation frequencies. They found that the wavelengths 
and growth rates of the most unstable KH modes differed considerably from the 
adiabatic case and that the nature of the cooling function, i.e.\ whether it 
is a steep or shallow function of temperature about its equilibrium value, 
along with the form of the heating function, is a crucial factor in 
determining initial growth rates. Their analysis was followed (Stone 
et al.\ 1997) by a time-dependent hydrodynamical simulation of a cooled jet 
that traced the growth of the KH instability in the non-linear regime. Stone 
et al.\ (1997) concentrated on the development of asymmetric modes. 
Such modes, and in particular the helical mode, might be responsible for the 
``wiggling'' motion seen in a number of YSO jets, e.g.\ HH~30 (Lop\'ez et al.\
1995) and may also transport momentum from the jet to its environment 
(Stone et al.\ 1997). Here we present the results of time-dependent 
hydrodynamical simulations of the growth of the KH instability in a radiatively 
cooled flow using parameters typical of YSO jets.  Rossi et al.\ (1997) 
reported the results of simulations of these axisymmetric modes of the KH 
instability in cylindrical symmetry and found that cooling appears to dampen 
the growth of the instability.  Theory suggests that the linear behaviour
of these modes will be similar in both slab and cylindrical symmetry.  Here
we present the results of simulations in slab symmetry and we compare our
results with the results of Rossi et al.\ (1997).

\section{Numerical method}

The code used is temporally and spatially second order accurate and takes
account of cooling due to ionisation and radiative atomic transitions.  It 
uses a MUSCL-type scheme (e.g. van Leer 1977) to integrate the 
hydrodynamic and ionisation equations.  The advection terms are calculated
in a straightforward upwind fashion, while all other differences are
centered.  The details and tests of this code are fully described in 
Downes (1996).  The simulations are purely hydrodynamic and are performed 
in slab symmetry.

\subsection{Equations}

The equations solved are
\begin{eqnarray}
\p{\rho}{t} & = & - \vec{\nabla} \cdot (\rho \vec{u}) \label{contin}\\
\p{\left(\rho \vec{u} \right)}{t} & = & - \vec{\nabla}[ \rho \vec{u}\otimes
\vec{u}+ \left(P+Q\right)\vec{I} ] \label{mom}\\
\p{e}{t} & = & - \vec{\nabla}
\cdot\left[\left(e+P+Q\right)\vec{u}\right]-L \label{energy} \\
\p{n_{\rm H} x}{t} & = & -\vec{\nabla}\cdot\left[n_{\rm H} x \vec{u}\right]
+J(x,n_{\rm H},T)
\label{ifrac} \\
\p{\rho \tau}{t} & = & - \vec{\nabla} \cdot (\rho \tau \vec{u}) \label{trace}
\end{eqnarray}
\noindent where $\rho$ is the mass density, $\vec{u}$ is the fluid
velocity, $P$ is the pressure, $Q$ is the artificial viscosity (following
von Neumann \& Richtmyer 1950), $I$ is the identity tensor, $e$ is the 
total energy density, $L$ is 
the energy loss, $n_{\rm H}$ is the number density of hydrogen atoms, $x$ is the
ionisation fraction of hydrogen, $J(x,n_{\rm H},T)$ is the rate of ionisation
of atomic hydrogen, and $\tau$ is a scalar introduced to track the jet 
material.   Equations \ref{contin} to \ref{energy} are the conservation 
equations for mass, momentum and energy, respectively.  Equation \ref{ifrac}
describes the ionisation fraction of hydrogen, and Eq. \ref{trace} is an
equation for $\tau$ which acts as a tracer for the jet material.

We assume the gas law to be
\begin{equation}
P=nkT
\end{equation}
where $n$ is the total number density of the gas and $k$ is Boltzmann's
constant.  The total energy density is then given by the equation
\begin{equation}
e=\frac{1}{2}\rho\vec{u}\cdot\vec{u} + \frac{c_v}{k}P
\end{equation}
\noindent where $c_v$ is the heat capacity of the gas.  The energy loss, 
$L$, is given by
\begin{equation}
L=L_{\rm rad}+E_0 J(x, n_{\rm H}, T)-H
\end{equation}
where $L_{\rm rad}$ is the energy loss due to radiative atomic transitions,
$E_0$ is the ionisation potential of atomic hydrogen, and $H$ is the
heating term used to maintain equilibrium at the initial temperature on the
grid.

The rate of ionisation of atomic hydrogen $J$ is given by
\begin{equation}
J(x,n_{\rm H},T)=n_{\rm H}^2\left\{x (1-x) c(T) - x^2 r(T)\right\}
\end{equation}
Here $c(T)$ and $r(T)$ are the collisional ionisation and radiative
recombination coefficients in cm$^3$ s$^{-1}$ respectively and are given by
\begin{eqnarray}
c(T) & = & 6.417\times10^{-11} T^{1/2} \exp\left(\frac{-1.579\times10^5}{T}
  \right) \\
r(T) & = & 2.871\times10^{-10} T^{-0.7}
\end{eqnarray}
(Falle \& Raga 1995).

\subsection{Scheme}

As an example of how we integrate Eqs. \ref{contin} to \ref{trace} let
us consider a continuity equation for an arbitrary variable $\psi$.  The scheme
we use is a two-step scheme which steps to time $(k+1/2)\delta t$ from 
$k\delta t$ using an upwind first order scheme.  The values of the variables
at time $(k+1/2)\delta t$ are then used to calculate fluxes which are
second order accurate in space.  These fluxes are used to integrate from 
time $k\delta t$ to time $(k+1)\delta t$.  This process leads to a scheme 
which is second order in time.  Specifically, we define 
$^1{\cal F}^k_{i+1/2}$, the first order flux across the cell 
boundary $(i+1/2)\delta x$ at time $k\delta t$, as 
\begin{equation}
^1{\cal F}^k_{i+1/2}=u^k_{i+1/2}a^k_{i+1/2}
\end{equation}
where
\begin{equation}
a^k_{i+1/2}=\up{\psi^k_i}{u^k_{i+1/2} }{\psi^k_{i+1}} 
\end{equation}
\noindent and $u^k_{i+1/2}$ is the advection velocity.  To calculate the
second order flux at time $(k+1/2)\delta t$ we make an estimate of the
gradient, $\frac{g^{\psi}_i}{\delta x}$, of the variable $\psi$ within cell 
$i$.  We do this using the equation 
\begin{equation}
g^{k,\psi}_i=\av{\psi^k_{i+1}-\psi^k_i}{\psi^k_i-\psi^k_{i-1}}
\end{equation}
where the function av is a non-linear averaging function given by (van Leer
1977)
\begin{equation}
\av{a}{b}=\left\{ \begin{array}{ll}
					\frac{2ab}{a+b} & \mbox{if $ab>0$} \\
					0 & \mbox{otherwise}
					\end{array} \right.
\end{equation}
We can now define the second order flux at $(i+1/2)\delta x$ and time
$k\delta t$ as 
\begin{equation}
^2{\cal F}^k_{i+1/2}=\left\{\begin{array}{ll}
									u^k_{i+1/2}[\psi^k_i+\frac{1}{2}g^{k,\psi}_i] &
										\mbox{if $u^k_{i+1/2}>0$} \\
									u^k_{i+1/2}[\psi^k_{i+1}-\frac{1}{2}g^{k,\psi}_{i+1}]
									& \mbox{otherwise}
									\end{array}
									\right.
\end{equation}
Note that the advection velocity $u_{i+1/2}$ itself cannot easily be 
calculated in an upwind fashion and is approximated by 
\begin{equation}
u^k_{i+1/2}=\frac{1}{2}\left\{[u^k_i+\frac{1}{2}g^{k,u}_i]+[u^k_{i+1}-
\frac{1}{2} g^{k,u}_{i+1}]\right\}
\end{equation}

A temporally and spatially second order accurate scheme is then given by 
first solving 
\begin{equation}
\psi^{k+1/2}_{i}=\psi^k_i-\frac{\lambda}{2}\left[^1{\cal F}^k_{i+1/2}-
{^1\cal F}^k_{i-1/2}\right]
\end{equation}
and then, using the values $\psi^{k+1/2}_{i}$ to calculate $^2{\cal
F}^{k+1/2}_{i\pm1/2}$, we solve the equation
\begin{equation}
\psi^{k+1}_{i}=\psi^k_{i}-\lambda\left[^2{\cal F}^{k+1/2}_{i+1/2}-
{^2\cal F}^{k+1/2}_{i-1/2}\right]
\end{equation}

\subsection{The cooling and heating functions}
\label{coo_heat}
\begin{figure}
\epsfxsize=8.8cm \epsfbox{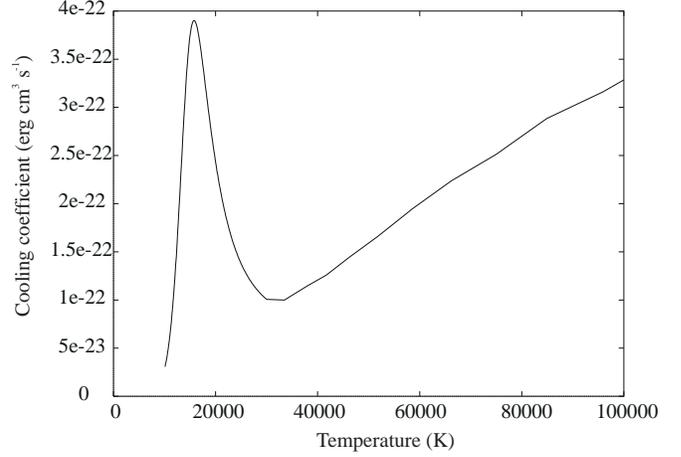}
\caption[]{\label{cooling}  Plot of the basic cooling function used in these
simulations.  This is a mixture of two functions - that used by Rossi et
al.\ (1997) for $T\leq3\times10^4$K and that given by Sutherland \& Dopita
(1993) otherwise}
\end{figure}

The cooling function used in this work is a combination of that used by
Rossi et al.\ (1997) and that contained in Sutherland \& Dopita (1993).  At low
temperatures ($T\leq3\times10^4K$) it is probable that the former is more
accurate.  However at higher temperatures this cooling function fails as it
only calculates cooling due to at most singly ionised atomic species.
The latter cooling function is calculated for a gas of cosmic abundances
cooling from about $3\times10^6$K and it does not take account of
non-equilibrium ionisation (unlike the Rossi et al.\ (1997) cooling 
function).  The resulting function is plotted in Fig. 
\ref{cooling}.  Tests of our implementation of the cooling have shown that 
it performs very well, predicting reasonably accurately the stability limit
as well as the amplitude of oscillation of an overstable radiative shock 
wave (see Downes 1996).  Note that this test is for rather strong shocks
($v_s\sim 100$ km s$^{-1}$) and so only tests the Sutherland \& Dopita
(1993) cooling function.

Rossi et al.\ (1997) also used a constant volume heating function which
cancels the cooling function at the equilibrium temperature, $T_{\rm eq}$.  
Here we 
compare the results of simulations which use this heating term with those
which simply assume that, below $T_{\rm eq}$ ($10^4$K in this
work) the cooling is insignificant and therefore can be set to zero.  Rather
than simply cutting off the cooling at $T_{\rm eq}$ we attenuate the cooling
function shown in Fig. \ref{cooling} as follows:
\begin{equation}
\label{cooling:tanh}
L_{\rm mod}(T)=L_{\rm rad}(T)\tanh\left[\left(\frac{T}{12\,000}\right)^{30}\right]
\end{equation}
This effectively implies that the cooling is zero at $10^4$ K.  Note 
that this means that the cooling function is a very
shallow function of temperature around $T_{\rm eq}$.  The assumption of
insignificant cooling below $T_{\rm eq}$ is often used in simulations of purely
atomic YSO jets as, below $10^4$K, the atomic cooling rate falls off 
rapidly (Blondin et al.\ 1990).  We find that these two approaches to 
maintaining
equilibrium in the initial configuration produce significantly different
results.  In addition we have run simulations where the heating term is
proportional to the density, as might be the case if radiative transfer 
was occurring.

\section{Initial and boundary conditions}

There are two approaches to analysing the KH instability.  One is to
perturb a flow at a particular point in space and observe how the
disturbance grows as it propagates downstream.  The other is to perturb a flow
everywhere and observe how the disturbance grows with time.  The former is
referred to as a spatial approach and has been adopted by, for example,
Hardee \& Norman (1988) and Hardee \& Stone (1997).  The latter is called
the temporal approach and has been used by Bodo et al.\ (1994) and Rossi et
al.\ (1997).  Here we use the temporal approach.

\begin{figure}
\epsfxsize=8.8cm \epsfbox{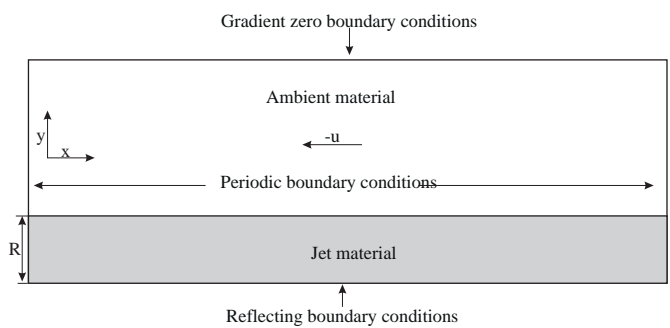}
\caption[]{\label{initial} The initial conditions used in the simulations 
of the KH instability}
\end{figure}

In order to do this the initial conditions were set up as shown in Fig.
\ref{initial}.  The slab jet occupies all of the grid within 1 jet radius of 
the `lower' edge.  The boundary conditions at either end of the grid are 
set to periodic so that, effectively, we have a jet with infinite length.  
The boundary away from the jet axis is set to have gradient zero boundary
conditions, while the jet axis boundary conditions are set to reflecting.

The grid cells are uniform in the direction parallel to the jet
axis with a size of $6\times10^{14}$ cm.  Perpendicular to this direction
the cell spacing obeys
\begin{equation}
\Delta y_j=\left\{\begin{array}{ll}
						\delta y & \mbox{if $j\leq130$} \\
						1.1\Delta y_{j-1} & \mbox{otherwise}
						\end{array} \right.
\end{equation}
\noindent where $j$ is the cell index perpendicular to the jet axis and
$\delta y=1\times10^{14}$ cm.  The grid size was set to 400$\times$200 cells,
or about 48$\times180$ jet radii.  We `stretch' the cells perpendicular to
the jet axis in order to avoid reflections from the upper boundary.  The jet
radius was set at 50 cells (i.e. $5\times10^{15}$ cm) which is typical of
YSO jets (e.g.\ Raga 1991).  

The jet to ambient density ratio was set to 1 and the jet and ambient
medium were set to be in pressure equilibrium also.  Three different density 
regimes were used: 20 cm$^{-3}$, 100 cm$^{-3}$ and 300 cm$^{-3}$.  These 
densities, although perhaps rather low for YSO jets (e.g.\ Bacciotti et al.\
1995), were selected to ensure that the cooling lengths behind
shocks were resolved. The pressure was chosen so that the initial 
temperature on the grid was $10^4$K.  This temperature being set as the
equilibrium temperature of the system (see Sect. \ref{coo_heat}).

The calculation was performed in the rest frame of the jet in order to
reduce the effect of advection errors on the growth of the instability.
The ambient medium was given an initial velocity of Mach 10 ($\sim$112 km
s$^{-1}$).  The jet material was given a small transverse velocity
perturbation:
\begin{equation}
v_y=\frac{0.07c}{N}\sin(\frac{\pi}{R} y)\sum_{j=1}^N \sin(k_j x)
\end{equation}
where $N=7$ is the number of perturbation wavelengths, $c$ is the jet sound
speed, $R$ is the jet radius, and $k_jR$ are approximately 0.21, 0.42, 0.52, 
0.84, 1.05, 1.68, and 2.09. These values were chosen so that the grid
length corresponded to an integral number of wavelengths of each 
perturbation.  The sine term outside the summation gives the perturbation a 
profile across the jet radius such that it is zero both on the axis and at 
the edge of the jet.  Therefore, the perturbation has a velocity structure
similar to that which would be produced by the first reflecting mode at all 
these values of $kR$.  However, it is not true to say that this
perturbation excites only the first reflecting mode since we do not
introduce corresponding perturbations in the pressure and density.

It is well known (e.g. Bodo et al.\ 1994) that introducing a shear layer
dampens the growth rates of all surface modes and body modes which have 
wavelengths shorter than the width of the layer.  In simulations of adiabatic
jets with high Mach numbers this is generally unimportant as the growth rates
of the surface waves are usually much less than the growth rates of the body 
modes.  However, in radiatively cooled jets analytic studies suggest that a 
surface mode exists which has growth rates higher than all other modes for a
large range of wavelengths if the cooling function around $T_{\rm eq}$ is
shallow enough (see Hardee \& Stone 1997).  Therefore it is important that 
we run simulations with different thicknesses for the shear layer in 
order to study the effect of this surface mode if it is present.

The jet itself is given a `top-hat' velocity profile modified by a term so 
that a shear layer exists initially between the jet and ambient medium.  
Two forms of this term have been tested.  The first is identical to that of 
Rossi et al.\ (1997), i.e.
\begin{equation}
u(y)=\frac{u_{\rm jet}}{\cosh\left[\left(\frac{y}{R}\right)^8\right]}
\end{equation}
which gives a shear layer of approximately 20 cells.  The second was chosen
to be
\begin{equation}
u(y)=u_{\rm jet}\left(1-\tanh\left[\left(\frac{y}{R}\right)^{30}\right]\right)
\end{equation}
This gives a shear layer of about 5 cells.  This latter shear layer is wide
enough
to be well resolved by the code, and hence avoids numerical errors which
lead to spurious waves being emitted from the jet boundary.  The latter was 
chosen for the majority of simulations in the parameter space study. 

\section{Analysis}

The analysis of the results of these simulations to find the growth rates
of the KH modes is not trivial.  First the velocity perpendicular to the
jet axis was averaged over 1 jet radius at a particular time.  A Fourier 
transform of the resulting data was performed and the power 
contained in the wavelengths of the initial perturbation were recorded.
This was done at several times during the simulation and these powers were
used to derive growth rates.

The averaging process described above was carried out to reduce the effect
of the different distributions of amplitude of the KH modes across the jet
radius.  The signal at a particular wavelength from a single row of cells
parallel to the jet axis will change with distance from the axis.  This 
signal will be given by a complicated weighted mean of the modes present.
The weighting associated with a particular mode is determined by the 
distribution of the disturbance due to that mode across the jet.  Thus, for
example, if we take a row of cells located very close to the edge of the
jet, the signal we receive might come primarily from surface modes.  If we
locate the row of cells a distance of $\frac{R}{2}$ from the jet axis the 
signal we will receive will be biased towards the first body mode, although
it may be modified significantly by the presence of other odd index body
modes.  Averaging reduces the effects of these problems.

It is important to note that the averaging process described above has the
effect of removing the signal from even-index body modes and it also
reduces the signal from surface modes.  The growth rates calculated on the 
basis of this averaging will be a weighted mean of the odd-index modes 
present at the wavelength in question, with the lower index modes being 
more heavily weighted.

Several tests were performed using this averaging technique for the growth
of the KH instability in an adiabatic jet with a Mach number of 10 and
density ratio of 1.  The
derived growth rates were within 20\% of those predicted by linear analysis 
where only one mode was present.  If more than one mode was present,
the derived growth rates at a particular wavelength lay between the growth 
rates predicted from theory for the modes present at that wavelength, as
expected.

\section{Results}

We have run 9 simulations with identical initial conditions, but with
different techniques for calculating the energy loss function $L$.
None of the simulations shown here have linear growth rates which are
significantly different to the adiabatic growth rates.  This is in
agreement with the linear results of Rossi et al.\ (1997), and would be 
expected anyway since the cooling times near equilibrium are much longer than
any other time scales in the system.  In addition we have run simulations,
both cooled and adiabatic, to test the effect of varying the shear layers as
described above.  It was found, in general, that widening the shear layer, 
while changing the behaviour of the system for a short time at the
beginning of the non-linear phase, did not significantly alter the long-term
evolution of the jet.

We will begin by describing the evolution of a typical radiatively cooled
simulation and comparing this with the evolution of an adiabatic simulation
with identical initial conditions.  We will then discuss the results of the
parameter space study in terms of 
\begin{itemize}
\item the transfer of momentum from jet material to ambient material
\item the distribution of momentum throughout the grid
\item the strength of shocks produced on the jet axis
\end{itemize}
Each of these sections will be broken down into an analysis of the effect
of the shear layer and an analysis of the effect of the different heating
terms.  All times are quoted in units of the sound crossing time
$\tau_{\rm c}=\frac{R}{c}$.  In the simulations presented here 
$\tau_{\rm c}\approx140$ yrs.

\subsection{General properties of the cooled KH instability}
\label{results:general}

As already noted, the linear behaviour of the KH instability was not
significantly altered by the introduction of cooling for the parameters 
chosen here.  This would be expected since the cooling time around 
equilibrium is longer than any other time-scales in these simulations.  

\begin{figure*}
\epsfxsize=14cm \epsfbox{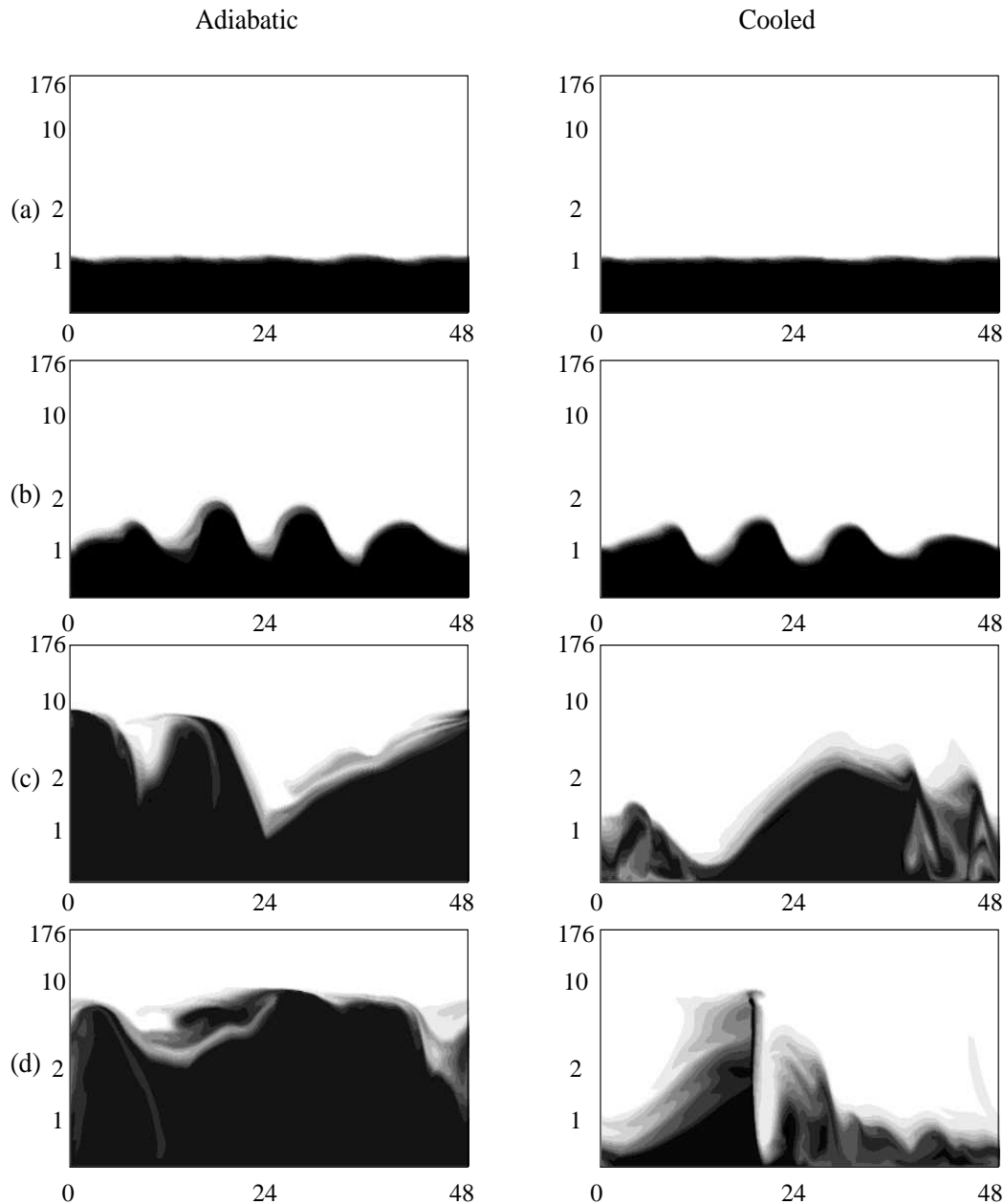}
\caption{\label{tracer} Grey-scale plots of the distribution of the jet 
tracer at time 8.5, 14.17, 19.12 and 25.51 labeled by {\bf a}, {\bf b},
{\bf c}, and {\bf d} 
respectively.  Black indicates pure jet material, while white indicates a
total absence of jet material.  The scale is linear.  Both simulations are
for jets of initial density 100 cm$^{-3}$.  Note how the jet material 
stays closer to the axis if the jet is cooled, and also that ambient 
material can be found close to the jet axis.  This is clearly not the case 
for the adiabatic simulation}
\end{figure*}

\begin{figure*}
\epsfxsize=14cm \epsfbox{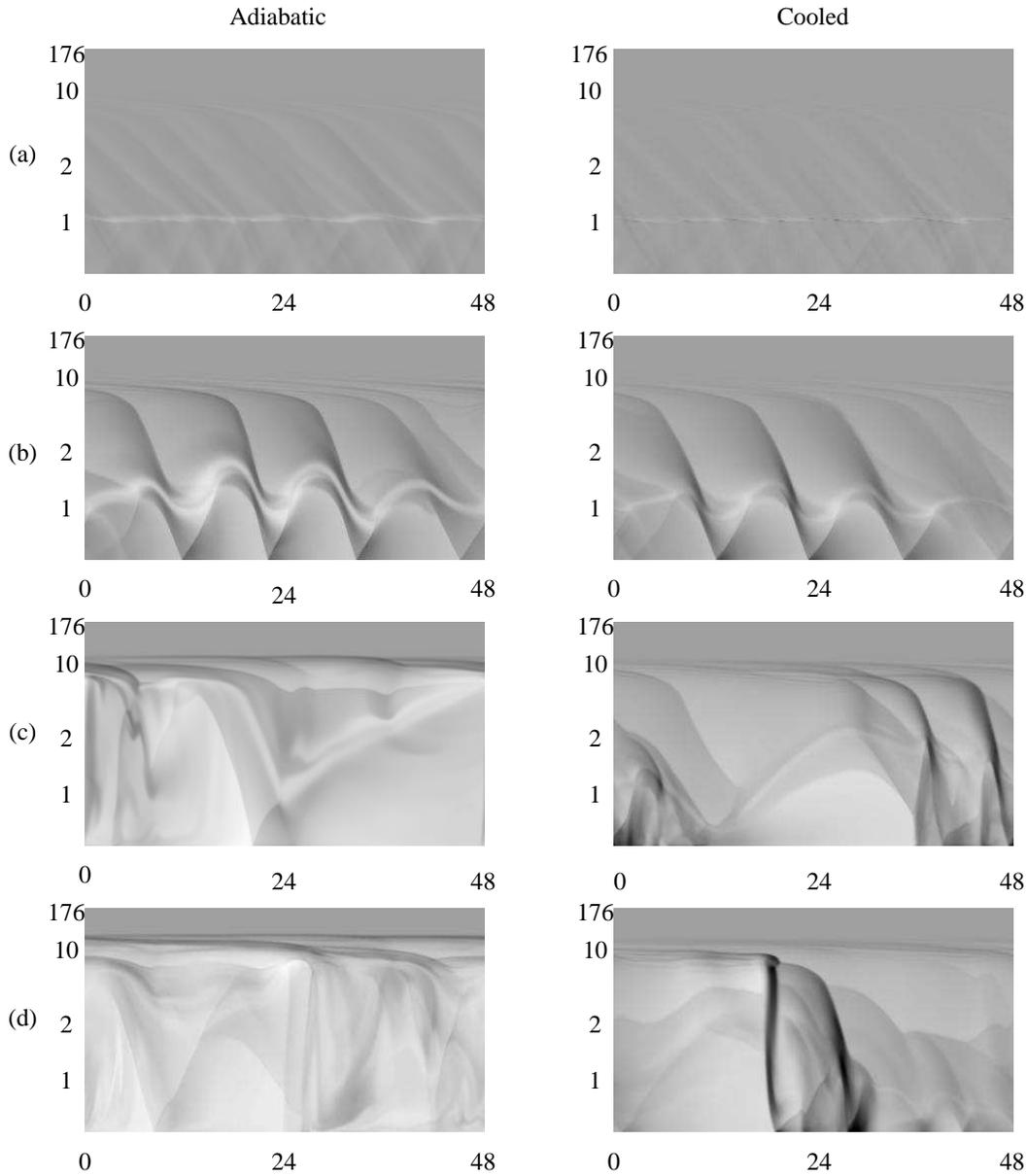}
\caption{\label{density} Grey-scale plots of the distribution of the
density for the same simulations, and at the same times, as those shown in Fig.
3.  A square-root scale is used ranging from 4 cm$^{-3}$ (white) to 700
cm$^{-3}$ (black)}
\end{figure*}

Once we enter the non-linear regime, however, cooling has a dramatic effect 
on the evolution of the system.  This is illustrated in Fig. \ref{tracer}
which contains grey-scale plots of the distribution of the jet tracer 
variable at various times throughout an adiabatic and a cooled simulation.
We can see that very little mixing occurs in the adiabatic simulation and
the jet expands due to the conversion of bulk kinetic energy to internal energy
by the growth of waves due to the KH instability.  Differences between the
adiabatic and cooled jets become noticeable at $t\approx14$ and by
$t=19$ there are very significant differences in the distribution of jet
material.  The cooled jet material remains closer to the axis and, in
addition, ambient material has been funneled onto the axis by the
distortion on the surface of the jet.  From this figure we would
intuitively expect stronger shocks to form in the cooled jet as a result of
this funneling of ambient material towards the axis.  In fact this is not
the case, due to the oblique nature of the shocks formed and also the damping
of the body modes by cooling observed by Rossi et al.\ (1997) and predicted 
by Hardee \& Stone (1997).  We can clearly see the nature of the shocks
formed in Fig.\ \ref{density} which contains plots of the density for the
same simulations at the same times as those shown in Fig. \ref{tracer}.

\begin{figure}
\epsfxsize=8.8cm \epsfbox{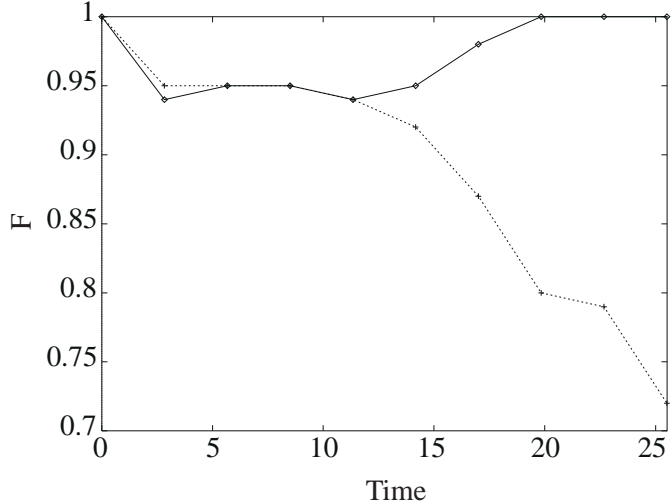}
\caption{\label{mixing} Plots of the proportion, $F$, of material which is jet 
material remaining within a distance $R$ of the axis.  The solid line denotes
the adiabatic jet and the dotted line denotes the cooled jet.  Note how
little jet material there is close to the axis in the cooled, compared to the 
adiabatic simulation}
\end{figure}

Figure \ref{mixing} shows the above results in a quantitative fashion.  It
contains plots of the proportion of material, $F$, lying between $y=0$ and 
$y=R$ which is jet material.  We can see that in the adiabatic case,
although a lot of jet material has moved away from the axis, almost all the
material remaining near the axis originated in the jet.  In contrast, if the
jet is allowed to cool radiatively, we can see that only 70\% of the material
lying between $y=0$ and $y=R$ is jet material by the end of the simulation
($t\approx25$).

\subsection{Transfer of momentum from jet to ambient material}
\label{mom_trans}
\begin{figure}
\epsfxsize=8.8cm \epsfbox{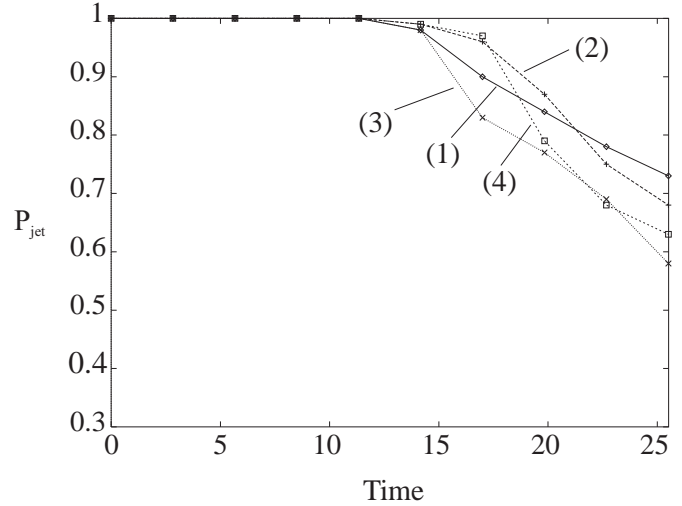}
\caption[]{\label{shear1} Plots of the fraction of momentum remaining
within jet material against time.  The plots are labeled as follows: (1)
and (2) denote the adiabatic jets with narrow and wide shear layers
respectively; (3) and (4) denote the cooled jets with narrow and wide shear
layers respectively. The results for the wide and narrow shear layers for
both the adiabatic and cooled simulations are very similar} 
\end{figure}

\begin{figure}
\epsfxsize=8.8cm \epsfbox{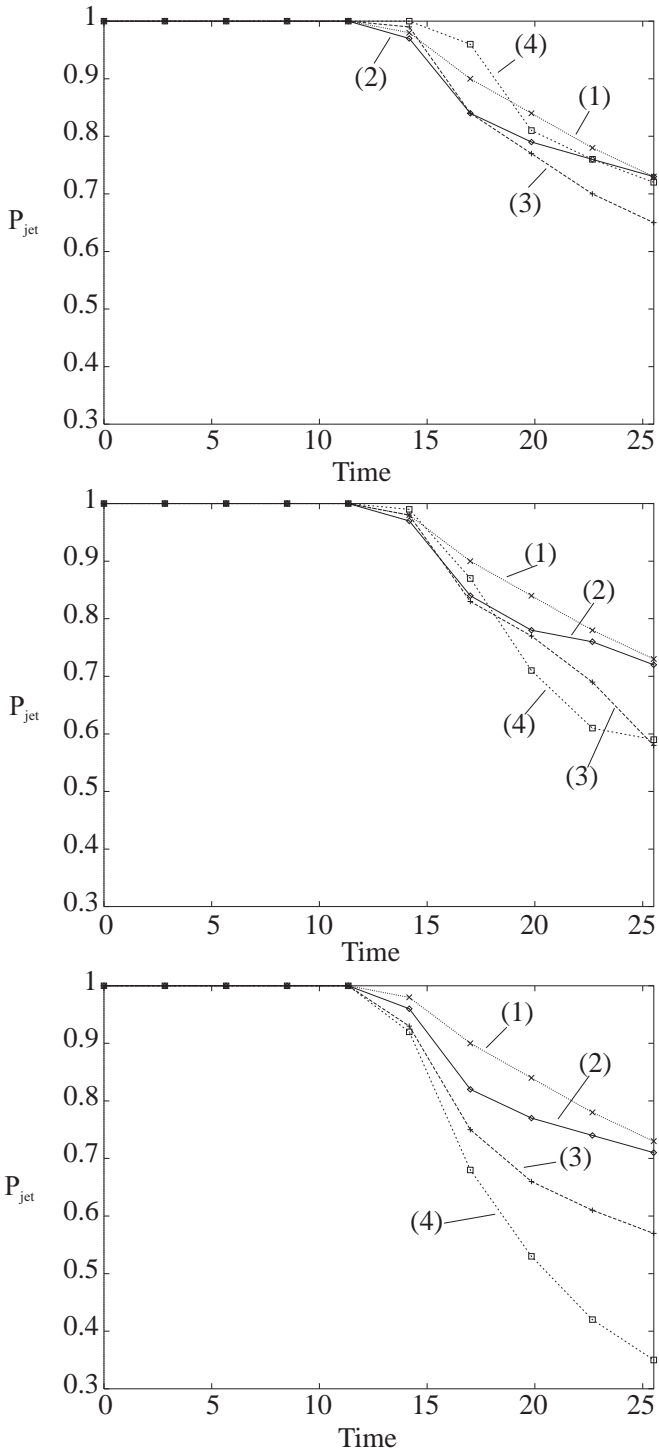}
\caption[]{\label{diff_heat} Plots of the fraction of momentum remaining
within jet material against time for simulations in which initial
equilibrium was maintained by a constant volume heating term, a heating term
proportional to $\rho$, and by assuming insignificant cooling below
$T_{\rm eq}$ (top to bottom).  The plots are labeled as follows:(1) denotes
the adiabatic jet result; (2), (3) and (4) denote simulations with
jet densities of 20 cm$^{-3}$, 100 cm$^{-3}$, 300 cm$^{-3}$ respectively}
\end{figure}

In this section we look at momentum transfer from material which was
initially in the jet to material which was initially in the ambient medium.
Note that this does not tell us about the spatial distribution of the momentum.
The momentum remaining in jet material at time $t$ is defined here as
\begin{equation}
P_{\rm jet}(t)=\sum_{\rm grid}\tau (t) u (t) \rho (t)
\end{equation}
In the plots shown in Figs. \ref{shear1} and \ref{diff_heat} this value 
is normalised by $P_{\rm jet}(0)$.

Two adiabatic simulations were run each with an initial shear layer with a
different width and it was found that, as expected, the results were not 
significantly different.  In particular, the momentum transferred to the 
ambient medium differed by less than 6\% throughout the entire duration of 
the simulation.  Varying the width of the shear layer has a larger effect 
on the radiative jets at $t\approx17$, as we can see in Fig.
\ref{shear1}.   The jet with the wide shear layer is initially more stable
than that with the narrow shear layer.  However, as noted above, the
state of the two jets seems to be rather similar by $t=20$.

Now we discuss the results of simulations with the `narrow' shear layer 
(described in Sect. \ref{coo_heat}) with different techniques of
maintaining initial cooling equilibrium.  The 
cooling function itself is the same for all the simulations.  We have run 9 
more simulations corresponding to 3 different heating terms and 3 different 
densities, each with a jet to ambient density ratio of 1.  

Figure \ref{diff_heat} contains plots of the fraction of momentum remaining 
in jet material against time for each of the 3 different densities and each
of the 3 techniques of maintaining equilibrium.  Each plot also shows the
behaviour of an adiabatic jet for comparison.  It is clear that in all
cases the radiative jets transfer momentum to the ambient material more
efficiently than the adiabatic jets.  It can also be seen that, for all
densities, assuming insignificant cooling below $T_{\rm eq}$ causes the most
efficient transfer of momentum.  If a heating term proportional to the
density is introduced the rate of transfer of momentum is reduced slightly.
If the heating term is a constant (i.e. constant volume) then the momentum
transfer is reduced more, though it still remains more efficient than
in the adiabatic case.

\subsection{Distribution of momentum}
\label{mom_dist}
Here we analyse how the momentum initially contained in the jet is
distributed throughout the grid with time. We define
\begin{equation}
P(y,t)=\frac{1}{N_y N_x}\sum_{j=0}^{N_y} \sum_{i=0}^{N_x} P_{i,j} (t)
\end{equation}
where $P_{i,j}(t)$ is the momentum at the grid point $(i,j)$ at time $t$ and
$N_x$ is the length of the grid in the along the axis of the jet in grid
cells and $N_y$ is chosen so that $\sum_{j=0}^{N_y}\Delta y_j=y$.  In the 
plots shown below we normalise these values by $P(R,0)$, i.e.\ the momentum 
contained within 1 jet radius of the axis at $t=0$.

We find that in both the adiabatic and the cooled
cases, widening the shear layer causes a noticeable difference in results
from around $t\approx15$ to $t\approx22$ with the jet with the narrow shear
layer distributing its momentum throughout the grid more quickly.  After
this time the differences in the initial widths of the shear layers has
almost no effect.

\begin{figure}
\epsfxsize=8.8cm \epsfbox{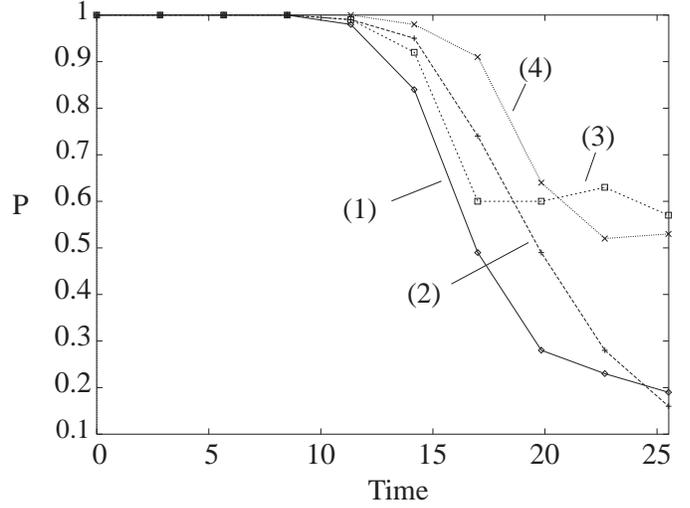}
\caption[]{\label{shear_dist} Plots of the fraction of momentum remaining
within 1 jet radius of the jet axis against time.  The labels (1), (2), (3)
and (4) denote the adiabatic result for narrow and wide shear layers and
the cooled results for narrow and wide shear layers, respectively.  Note
how the cooled jets reach a quasi-steady state by $t\approx20$ while the
adiabatic jets continue to lose momentum}
\end{figure}

\begin{figure}
\epsfxsize=8.8cm \epsfbox{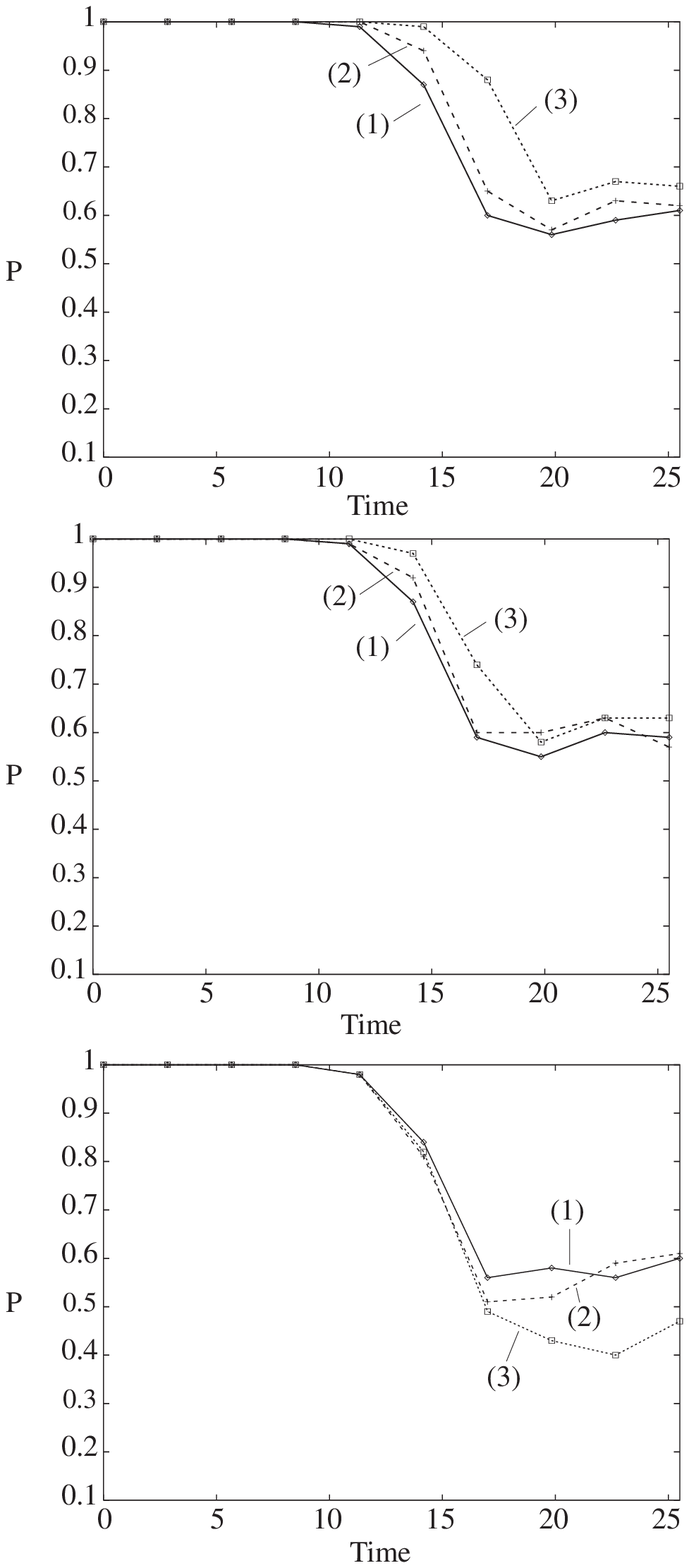}
\caption[]{\label{heat_dist} Plots of the fraction of momentum remaining
within $R$ of the jet radius against time in which initial equilibrium was
maintained by a constant volume heating term, a heating term proportional
to $\rho$, and by assuming insignificant cooling below $T_{\rm eq}$ (top to 
bottom).  The labels (1), (2) and (3) denote the simulations with densities
of 20 cm$^{-3}$, 100 cm$^{-3}$ and 300 cm$^{-3}$.  The top two plots
indicate that the loss of momentum decreases with increasing density while
the reverse is the case for the bottom plot}
\end{figure}

In Sect. \ref{mom_trans} we noted that radiative cooling caused a
more efficient transfer of momentum from jet to ambient material.  However,
it is clear from Fig. \ref{heat_dist} that this momentum remains closer to 
the jet axis if the jet is cooled.  

A particularly interesting result is that when the jet has entered the
non-linear regime the momentum distribution enters a quasi-steady state with
between 50\% and 70\% of the initial jet momentum remaining within 1 jet
radius of the axis.  This implies that the shocks formed during the growth
of the instability tend not to force longitudinal momentum `sideways' to
beyond a couple of jet radii.  It is also clear that increasing the density
reduces the rate of loss of momentum from this region for the simulations 
with a heating term while the reverse is true for the simulations assuming 
that cooling is zero below $T_{\rm eq}$.

From Fig. \ref{heat_dist} we can see that the momentum distribution 
attains this quasi-steady state at $t\approx20$ crossing times for all the 
heating terms and densities investigated here.  Generally, the simulations 
with a constant volume heating term retain a higher proportion of jet 
momentum in this region.  The simulations which used the assumption of 
insignificant cooling below $T_{\rm eq}$ lose most momentum from this region.

It is interesting to note that, for $t\leq17$, the simulation with
the assumption of insignificant cooling below $T_{\rm eq}$ is very similar to an
adiabatic system in terms of the momentum distribution.  However, the 
adiabatic system continues to transport momentum beyond $y=R$ after the 
cooled simulations have reached a quasi-steady state.

Overall, these results are quite surprising since the radiative
jets have been found (see Sect. \ref{mom_trans}) to transfer {\em more} 
momentum to ambient material. This is a reflection of the result that 
significant amounts of ambient material are transported to within 1 jet 
radius of the jet axis (see Sect. \ref{results:general}).

\subsection{Shock strengths and morphologies}

We have measured the maximum shock strength on the axis of the jet against
time.  Since shocks are typically smeared over 3 to 4 cells in this code we
measure the maximum shock strength using the equation
\begin{equation}
{\cal S}=\max_{\rm axis} |u_{i,j}-u_{i-4, j}|
\end{equation}
Thus we only look at the amplitude of the velocity discontinuity in the
direction parallel to the jet axis.

\begin{figure}
\epsfxsize=8.8cm \epsfbox{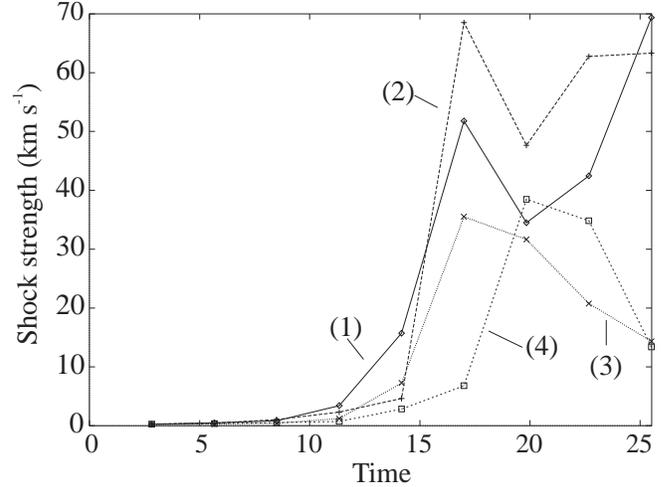}
\caption[]{\label{shear_shock} Plots of the amplitude of the largest velocity
discontinuity on the jet axis against time for the adiabatic and cooled
jets with narrow and wide shear layers. The labels (1), (2), (3) and (4)
denote the adiabatic results for narrow and wide shear layers, and the
cooled results for narrow and wide shear layers respectively.  See text} 
\end{figure}

Figure \ref{shear_shock} shows plots of the maximum shock strength against
time for the adiabatic and cooled simulations with the wide and narrow
shear layers.  We can see that shocks form earlier in the adiabatic jet
with the narrow shear layer and reach a maximum strength of about 50 km 
s$^{-1}$.  The jet with a wide shear layer takes slightly longer to develop
shocks but, when they do form, they reach a higher amplitude of almost 70 km
s$^{-1}$. 

The same type of effect of the shear layer is seen in the cooled case.
Here, though, the differences in the time taken for shocks to form is
greater while the difference in the maximum shock strength attained
throughout the simulations is less.

\begin{figure}
\epsfxsize=8.8cm \epsfbox{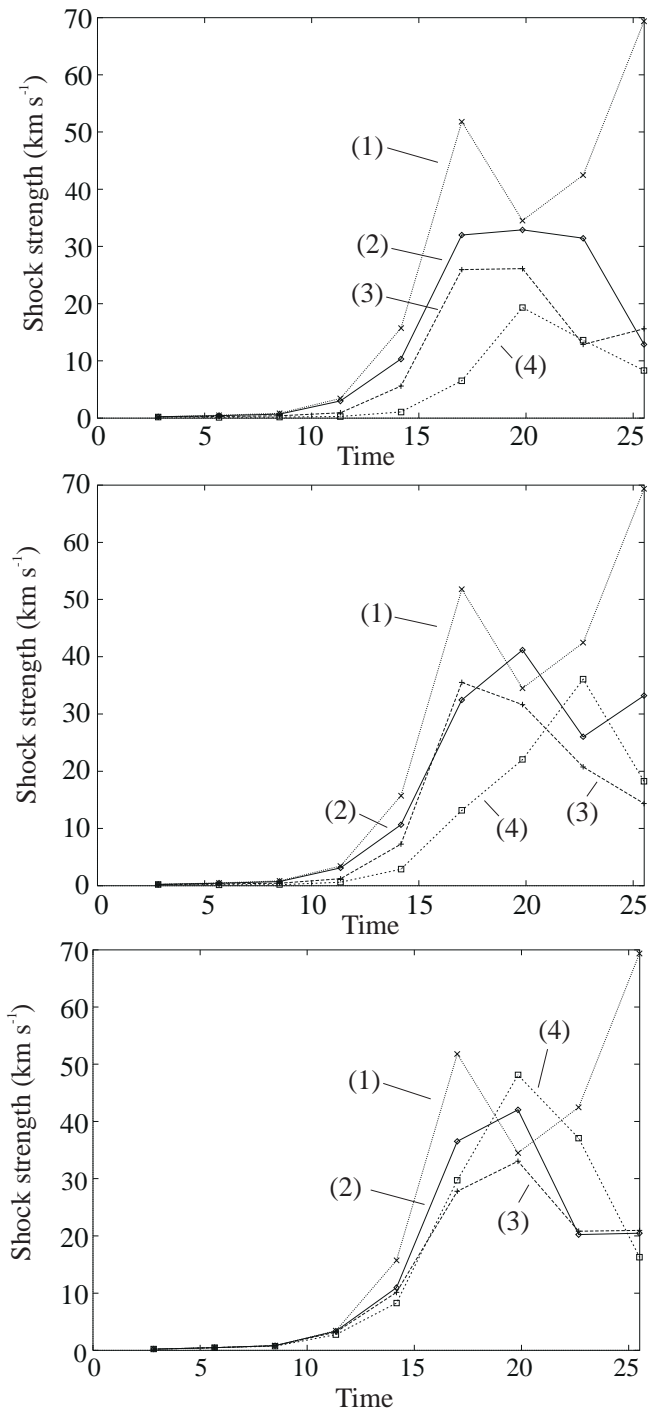}
\caption[]{\label{heat_shock} Plots of the amplitude of the largest
velocity discontinuity on the jet axis against time for simulations in 
which initial equilibrium was maintained by a constant volume heating term,
a heating term proportional to $\rho$, and by assuming insignificant cooling
below $T_{\rm eq}$ (top to bottom).  The labels (1), (2), (3) and (4) denote 
the adiabatic result, and simulations with densities of 20 cm$^{-3}$, 
100 cm$^{-3}$ and 300 cm$^{-3}$ respectively.  See text}
\end{figure}

Figure \ref{heat_shock} shows plots of the maximum shock strength against
time for the adiabatic simulation with the narrow shear layer and the 9
other simulations with different densities and heating terms.  In all cases
where the heating term is either constant or proportional to the density it 
is clear that increasing the density slows down the development of strong 
shocks and reduces the maximum strength of these shocks over the duration of
the simulation.  The same cannot be said, however, for the simulations 
where we assume that cooling is insignificant below $T_{\rm eq}$.  Here the
simulation with $\rho=300$ cm$^{-3}$ develops stronger shocks than either
of the simulations with $\rho=100$ cm$^{-3}$ or 20 cm$^{-3}$.

As might be expected from Sects. \ref{mom_trans} and \ref{mom_dist}, the
heating term which leads to the slowest and weakest shock development is
the constant volume one.  If the heating term is set proportional to $\rho$
the shocks are formed slightly faster and evolve to a greater strength.
The simulation which simply assumes the cooling to be insignificant below
$T_{\rm eq}$ forms the strongest shocks.

In general, the shocks produced by the instability in radiative jets are
quite flat, but with a slight bow shape pointing {\em towards} the jet source.  
In the adiabatic simulations these shocks tend to be more curved with the 
apex of the curve pointing away from the jet source.

\subsection{Proper motions}

The proper motions of the shocks produced in the simulations have been
measured as the motion of the shock with respect to the ambient medium.  The 
adiabatic simulations produced shocks with proper motions in the range 
0.7$u_{\rm jet}$--0.9$u_{\rm jet}$ while the shocks in the cooled simulations
moved with proper motions of 0.3$u_{\rm jet}$--0.9$u_{\rm jet}$.  No
significant differences were found between the proper motions of shocks
produced by different heating terms or densities.  Strong shocks were found
to move slower with respect to the jet source than weaker shocks, as would
be expected from momentum balance arguments.

While making these measurements it was noted that, in cooled jets, the 
shock pattern tends to coalesce into 1 shock within a few crossing times of 
the first shocks appearing.  The rate of coalescence was found to increase
with the density of the jet.  Shocks were not found to coalesce in the 
adiabatic simulations.

\section{Discussion}

We have performed a total of 12 different simulations of the KH instability 
in jets.  The parameters were chosen to approximate the physical conditions 
thought to prevail in jets from young stellar objects.  

\subsection{General differences introduced by cooling}
\label{general_differences}

We have found that introducing cooling
\begin{itemize}
\item increases the level of mixing between jet and ambient material
\item increases the amount of momentum transferred from jet material to
ambient material
\item increases the time taken for shocks to develop in the flow
\item reduces the strength of these shocks
\item reduces the rate of decollimation of the momentum flux
\end{itemize}
The first and second of these results are somewhat surprising as they 
contradict the results presented by Rossi et al.\ (1997).  However,
we found it possible to reproduce the results of Rossi et al.\ (1997) by
performing the simulations in cylindrical symmetry.  Therefore we can
conclude that, although in the linear regime we do not expect significant
differences between the 2 symmetries, {\em the non-linear growth of the 
instability is significantly affected by the choice of symmetry}.

We will now explain how radiative jets can transfer more momentum to the 
ambient medium while the momentum flux in these systems remains better 
collimated than in an adiabatic system.  As the KH instability grows, 
longitudinal momentum will be converted to internal energy by the resulting 
compressions and shocks within the jet.  If the jet is not cooled it will, as
a result, become overpressured and begin to expand and so momentum will be 
moved away from the jet axis.  This mechanism allows longitudinal momentum to
be transported away from the jet axis while still remaining in jet
material.  If, on the other hand, the jet is cooled this internal energy
is radiated away causing the jet to remain in approximate pressure 
equilibrium with the surrounding medium and thus momentum will not be
transported away from the jet axis in this way.  In fact, as the
instability grows in the cooled jet the surface of the jet develops a
saw-tooth profile which steepens and eventually funnels ambient material
towards the jet axis, causing mixing.  It is in this way that momentum is
transferred from jet to ambient material in the radiatively cooled
simulations.  It is interesting that the shocks produced by this process do
not appear to decollimate the jet.  

\begin{figure}
\epsfxsize=8.8cm \epsfbox{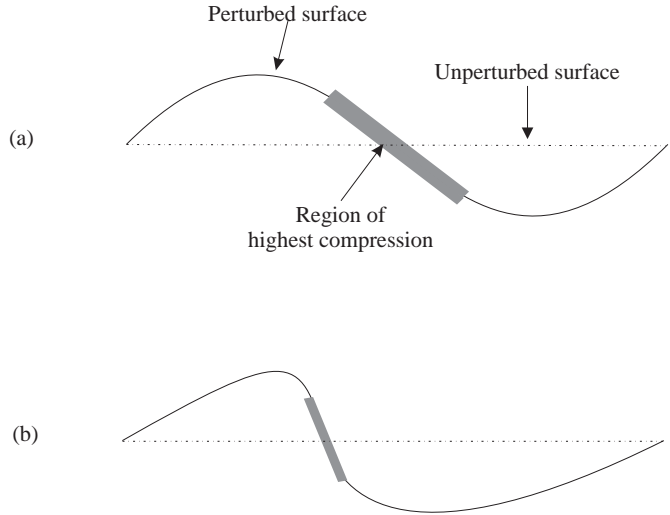}
\caption{\label{steepening}  Schematic diagram of how the cooling affects the 
growth of the perturbation on the surface of the jet in non-adiabatic 
simulations of the KH instability.  Initially, in (a),  the perturbation is 
sinusoidal (due to the sinusoidal nature of the velocity perturbation).  
This quickly develops into a saw-tooth pattern (as in (b)) due to the inertia 
of the high density material formed in the region of highest compression.  
See text}
\end{figure}

The process by which the steepening of the disturbance on the surface of 
the jet occurs can be understood as follows.   As the jet's surface becomes
perturbed by the transverse velocity perturbation, the expanding parts of 
the jet begin to compress the surrounding ambient medium.  This process 
heats both ambient and jet material near the boundary between the two.
The region is then cooled and a dense filament will form at the boundary.
The inertia of this filament will cause the smooth wave pattern to be
converted to a saw-tooth one as the wave grows in amplitude.  This process
is illustrated in Fig. \ref{steepening}.  Put another way, the waves
steepen and `break' earlier in the radiative than the adiabatic case because
of a stronger dependence of the wave speed on its amplitude.

The shocks produced in the adiabatic jet are formed by the growth of body
modes.  Our results suggest that cooling of the type expected in YSO jets 
inhibits the development of shocks and this is in agreement with 
the results of Rossi et al.\ (1997).  This is because the development of
the instability takes longer due to the increased levels of cooling in low
density regions and vice versa for high density regions with respect to
equilibrium values (see Hardee \& Stone 1997).  Therefore, in a cooled jet,
the time taken for shocks to develop in the jet is increased.  Note however,
that while this explanation is valid for these simulations, it does not
imply that, in general, shocks resulting from the KH instability are weaker
if cooling is introduced.  We emphasize that whether this is the case in
general depends on the form of the heating and cooling functions.

\subsection{Effect of shear layers}

Our simulations suggest that widening the shear layer does not
significantly affect the long term evolution of the system.

This result is easily explained.  Previous studies of the KH instability in
adiabatic jets (e.g.  Payne \& Cohn 1985; Hardee \& Norman 1988) have 
shown that the surface modes do not grow as fast as the body modes in high
Mach number flows for $kR\sim1$.  A wide shear layer tends to damp
the growth of surface modes and body modes with wavelengths less than the
width of the layer (for example, Bodo et al.\ 1994).  Therefore, although 
widening the shear layer will damp surface modes, we know that in an 
adiabatic jet these modes do not have a significant effect on the jet
anyway.  Since it is these modes which are expected to disrupt the jet and thus 
cause mixing and transfer of momentum from jet material to ambient 
material, we do not expect widening the shear layer to have much effect on
momentum transfer in the adiabatic case.  
According to Hardee \& Stone (1997) introduction of cooling can allow 
a new surface mode to grow which has higher growth rates than any other 
modes present if the cooling function is shallow enough around $T_{\rm eq}$.  
In our simulations it is difficult to tell whether the cooling function is 
shallow enough around $T_{\rm eq}$ as it is a non-equilibrium one and
we also include energy dumped into ionisation in our calculations.  Since
the differences in growth rates introduced by the narrowing of the shear 
layer are of order a few percent, it seems reasonable to say that this 
surface mode does not exist, or at least has a very low growth rate, for the
energy loss functions studied here.

\subsection{Effect of changing the heating function}

Detectable differences were found between simulations which included a
constant volume heating term and those which included a heating term
proportional to $\rho$.  Generally, the inclusion of the former resulted in
behaviour more similar to an adiabatic system than simulations with the
latter kind of heating term, though this difference is relatively small.

Simulations with the assumption that the cooling is insignificant below 
$T_{\rm eq}$ were dramatically different from adiabatic simulations and were
generally more unstable than the simulations which used a heating function
to maintain initial equilibrium.  This difference is unlikely to
arise from the growth of the surface mode mentioned by Hardee \& Stone
(1997) because of the very low level of the cooling around $T_{\rm eq}$ for
the modified cooling function described in Eq. \ref{cooling:tanh}.  It is
more likely to be due to the fact that there is no heating term in the 
system and this allows the instability to grow faster than in the other 
cooled systems.

The injection of energy by a heating term makes the system as a
whole more similar to an adiabatic one.  It is plausible to suggest that
the lack of a heating term will therefore accentuate any differences
between adiabatic and cooled systems.

\subsection{Differences between slab and cylindrical symmetry}

As already stated, the results of simulations of the cooled KH instability
in slab and cylindrical symmetry differ in 2 respects.  In slab symmetry
the introduction of cooling 
\begin{itemize}
\item increases the level of mixing between jet and ambient material
\item increases the amount of momentum transferred from jet material to
ambient material
\end{itemize}
whereas the introduction of cooling in cylindrical symmetry does the
opposite.

We can understand this as follows.  The increase in mixing and momentum
transfer in slab symmetry results from disturbances on the surface of the
jet steepening due to cooling as discussed in Sect.
\ref{general_differences}.  While we do not expect this steepening process to
be significantly modified by the choice of symmetry we note that axisymmetric
waves of a given amplitude on the surface of the jet require higher pressures 
(and hence higher temperatures) on the jet axis in cylindrical symmetry.  This 
is because the amplitude of an axisymmetric wave in cylindrical symmetry goes 
as $r^{-\frac{1}{2}}$ whereas it is independent of $r$ in slab symmetry.
Thus if the system is cooled, and the cooling function is an increasing 
function of $T$, we expect cooling to damp the waves more in cylindrical 
than in slab symmetry.  Hence the disturbances on the surface of the jet will 
be lower in amplitude and take longer to steepen.  This is precisely 
what we observe in cylindrically symmetric simulations.  By the time the 
waves have steepened significantly the corresponding adiabatic jet will have
disrupted.

Hence we conclude that, while slab symmetric simulations are adequate to
approximate the linear behaviour of the axisymmetric modes of the KH 
instability in cooled cylindrical jets, such simulations give misleading
results in the the non-linear regime.  In principle one would expect the above 
argument to hold for studies which examine the growth of the helical modes of 
the KH instability using slab symmetric simulations of the sinusoidal modes 
(e.g.\ Stone et al.\ 1997).  However, this assumes that the most significant 
waves are ones which pass through the body of the jet rather than along its 
surface. 

\begin{acknowledgements} 
This work has been partly supported by Forbairt grant number BR/93/013/.  
We would also like to thank Alex Raga, Stephen O'Sullivan, Milena Micono,
Jim Stone, Silvano Massaglia and Phil Hardee for useful discussions during 
the preparation of this paper.
\end{acknowledgements}

\section{References}
\rf{Bacciotti, E., Chiuderi, C., Oliva, E., 1995, \AAP {296} {185}}
\rf{Bachiller, R., 1996, ARA\&A 34, 111}
\rf{Birkinshaw, M., 1991, in: Beams and Jets in Astrophysics, ed.\ P. Hughes,
(Cambridge: CUP), p.\ 278}
\rf{Blondin, J.M., Fryxell, B.A., K\"onigl, A., 1990, \APJ {360} {370}}
\rf{Bodo, G., Massaglia, S., Ferrari, A., Trussoni, E., 1994, \AAP {283} 
{655}} 
\rf{B\"uhrke, T., Mundt, R., Ray, T.P., 1988, \AAP {200} {99}}
\rf{Downes, T.P., 1996, PhD Thesis (University of Dublin, Ireland)}
\rf {Edwards, S.E., Ray, T.P., Mundt, R., 1993, in: Protostars and Planets 
III, eds.\ E.H.\ Levy J.I.\ Lunine, (Tucson: University of Arizona), p.\ 567}
\rf{Eisl\"offel, J., Davis, C.J., Ray, T.P., Mundt, R., 1994 \APJL {422}
{91}}
\rf{Falle, S.A.E.G., Raga, A.C., 1995, MNRAS 272, 785}
\rf{Harris, D.E., Biretta, J.A., Junor, W., 1997, \MN {284} {L21}} 
\rf{Hardee, P.E., Cooper, M.A., Clarke, D.A., 1994, \APJ {424} {126}}
\rf{Hardee, P.E., Stone, J.M., 1997, \APJ {483} {121}}
\rf{Hardee, P.E., Norman, M.L., 1988, \APJ {334} {70}}
\rf{Hartigan, P., Morse, J.A., Raymond, J., 1994, ApJ 436, 125}
\rf{Lop\'ez, R., Raga, A., Riera, A., Anglada, G., Estalella, R., 1995, 
\MN {274} {L19}}
\rf{Neckel, Th., Staude, H.J., 1987, \APJ {322} {L27}}
\rf{Padman, R., Bence, S., Richer, J., 1997, in: Herbig-Haro Outflows and 
the Birth of Low Mass Stars, IAU Symposium No.\ 182, eds.\ B.\ Reipurth \& 
C.\ Bertout, (Dordrecht: Kluwer Academic Publishers, p.\ 123}
\rf{Payne, D.G., Cohn, H., 1985, \APJ {291} {655}}
\rf{Raga, A.C., 1991, \AJ {101} {1472}}
\rf{Raga, A.C., Canto, J., Binette, L., Calvet, N., 1990, \APJ {364} {601}}
\rf{Ray, T.P., 1996, in: Proceedings of the NATO ASI on Solar and 
Astrophysical MHD Flows, ed.\ K.\ Tsinganos, (Dordrech: Kluwer Academic 
Publishers), p.\ 539}
\rf{Ray, T.P., Mundt, R., Dyson, J., Falle, S.A.E.G., Raga, A.,
1996, \APJ {468} {L103}}
\rf{Reipurth, B., Hartigan, P., Heathcote, S., Morse, J.A., Bally, J.,
1997, \AJ {114} {757}}
\rf{Rossi, P., Bodo, G., Massaglia, S., Ferrari, A., 1997, \AAP {321}
{672}}
\rf{Stone, J.M., Norman, M.L., 1993, \APJ {413} {210}}
\rf{Stone, J.M., Norman, M.L., 1994, \APJ {420} {237}}
\rf{Stone, J.M., Xu, J., Hardee, P.E., 1997, \APJ {483} {136}}
\rf{Sutherland, R.S., Dopita, M.A., 1993, \APJS {88} {253}}
\rf{van Leer, B., 1977, J. Comp. Phys. 23, 276}
\rf{von Neumann, J., Richtmyer, R.D., 1950, J. Appl. Phys. 21, 232}
\end{document}